\documentclass{PoS}%
\usepackage{graphics}
\usepackage{bm}
\usepackage{color}
\usepackage{fancybox}
\usepackage{amsmath}
\usepackage{rsfs}
\usepackage{comment}
\usepackage{cite}
\usepackage{amsfonts}
\usepackage{amssymb}
\usepackage{graphicx}%
\providecommand{\U}[1]{\protect\rule{.1in}{.1in}}
\title{%
\vspace{-4mm}
\begin{flushright}
\begin{minipage}{4cm}
\normalsize 
KEK Preprint 2019-47  \\ 
CHIBA-EP-242  \\
\end{minipage} 
\end{flushright}\vspace{3mm}
Quark confinement in the Yang-Mills theory 
\\
with a gauge-invariant gluon mass 
\\
in view of the gauge-invariant BEH mechanism %
}

\ShortTitle{Quark confinement in the Yang-Mills theory with a gauge-invariant gluon mass}

\author{\speaker{Akihiro Shibata}\\
 Computing Research Center, High Energy Accelerator Research Organization (KEK),   Tsukuba 305-0801, Japan ; \\
  SOKENDAI (The Graduate University for Advanced Studies), Tsukuba 305-0801, Japan\\
        E-mail: \email{akihiro.shibata@kek.jp}}

\author{Kei-Ichi Kondo\\
        Department of Physics, Graduate School of Science, Chiba University, Chiba 263-8522, Japan
\\
        E-mail: \email{kondok@faculty.chiba-u.jp}}

\author{Ryutaro Matsudo\\
        Department of Physics, Graduate School of Science and Engineering, Chiba University, Chiba 263-8522, Japan
\\
        E-mail: \email{afca3071@chiba-u.jp}}

\author{Shogo Nishino\\
        Department of Physics, Graduate School of Science, Chiba University, Chiba 263-8522, Japan
\\
        E-mail: \email{shogo.nishino@chiba-u.jp}}

\abstract{
In order to clarify the mechanism of quark confinement in the Yang-Mills theory with mass gap, we propose to investigate the massive Yang-Mills model, namely, Yang-Mills theory with ``a gauge-invariant gluon mass term'', which is to be deduced from a specific gauge-scalar model with a single radially-fixed scalar field under a suitable constraint called the reduction condition. %
The gluon mass term simulates the dynamically generated mass to be extracted in the low-energy effective theory of the Yang-Mills theory and plays the role of a new probe to study the phase structure and confinement mechanism.
In this talk, we first explain why such a gauge-scalar model is constructed without breaking the gauge symmetry through the gauge-independent description of the Brout-Englert-Higgs mechanism which does not rely on the spontaneous breaking of gauge symmetry. %
Then we discuss how the numerical simulations for the proposed massive Yang-Mills theory can be performed by taking into account the reduction condition in the complementary gauge-scalar model on a lattice. 
By using the reweighting method, we have investigated the effect of the gluon mass term to the Wilson loop (the static potential) and the dynamically generated mass. 
Moreover, we point out that the adjoint case would gives an alternative understanding for the physical meaning of the gauge-covariant decomposition for the Yang-Mills field known as the Cho-Duan-Ge-Faddeev-Niemi decomposition, while the fundamental case would give a novel decomposition which has been overlooked so far.

}

\FullConference{37th International Symposium on Lattice Field Theory - Lattice2019\\
		16-22 June 2019\\
		Wuhan, China}

\begin{document}
\section{Introduction}

We want to ask what is the \textit{mechanism of quark confinement}? A
promising scenario for quark confinement is the \textit{dual superconductor
picture} of the QCD vacuum proposed by Nambu, 't Hooft and Mandelstam
\cite{dualsuper},
which is supposed to be caused by condensation of magnetic objects, such as
magnetic monopoles, dyons, etc. One of the remarkable facts on this picture
found in the preceding studies is: \textit{Infrared Abelian dominance}: The
Abelian part (or diagonal component) of the gauge field becomes dominant for
quark confinement in the low-energy or long-distance region \cite{EI82}.
This hypothesis was confirmed by calculating the Wilson loop average on the
lattice. The string tension of the linear potential in the static
quark-antiquark potential can be reproduced by the Abelian part alone
\cite{SY90}.
[\textit{Abelian dominance of the string tension}] Another indication is that
the off-diagonal gluon propagator exhibits the exponential fall-off in the
distance \cite{AS99}.
[\textit{Abelian dominance of the diagonal propagator}] This suggests the
\textit{dynamical generation of the off-diagonal gluon mass}.

However, these results were obtained only in the specific gauge called the
\textit{maximal Abelian (MA) gauge} based on the idea of \textit{Abelian
projection method} proposed by 't Hooft \cite{tHooft81}.
Therefore, the gauge invariance or independence was not clear in these studies.

Recently, we have succeeded to demonstrate the \textit{Abelian dominance of
the string tension in the gauge-invariant way} based on the novel
reformulation of the Yang-Mills theory in terms of the new field variables
obtained from the \textit{gauge covariant decomposition} and the
\textit{non-Abelian Stokes theorem for the Wilson loop operator}. See a review
\cite{KKSS15} for more details.

How about the \textit{Abelian dominance of the diagonal propagator}? The
propagator can be obtained only after the gauge fixing. Therefore, Abelian
dominance of the diagonal propagator cannot be extended in the gauge invariant
way. 
Instead, however, we can give a gauge-invariant definition for the
\textit{off-diagonal gluon mass}. Therefore, we can study the mass generation
of the off-diagonal gluon mass in the gauge-invariant way. This is based on
the \textit{gauge-independent description of the Brout-Englert-Higgs (BEH)
mechanism} \cite{Higgs1} proposed recently by one of the authors
\cite{Kondo16},
which needs neither the \textit{spontaneous breaking of gauge symmetry} $G \to
H$, nor the \textit{non-vanishing vacuum expectation value of the scalar
field} $\langle0| \phi(x) |0 \rangle:= v\ne0$. To explain it, we need to
introduce a specific gauge-scalar model (\textit{complementary gauge-scalar
model}) which reduces to the \textit{Yang-Mills theory with a gauge-invariant
gluon mass term} (\textit{massive Yang-Mills theory}). The gauge-invariant
gluon mass term simulates the dynamically generated mass to be extracted in
the low-energy effective theory of the Yang-Mills theory and plays the role of
a new probe to study confinement mechanism through the phase structure
(Confinement, Higgs, deconfinement) in the gauge-invariant way. In this talk
we give preliminary studies in this direction.

\section{Gauge-independent BEH mechanism for the gauge-scalar model}

We consider $G=SU(2)$ gauge-scalar model with a single \textit{adjoint scalar
field} characterized by the gauge-invariant Lagrangian:
\begin{align}
\mathscr{L}_{\mathrm{GS}} = \mathscr{L}_{\mathrm{YM}} +
\mathscr{L}_{\mathrm{kin}} = - \frac{1}{2} \mathrm{tr} \{\mathscr{F}^{\mu\nu
}(x) \mathscr{F}_{\mu\nu}(x) \} + \mathrm{tr} \{ ( \mathscr{D}^{\mu
}[\mathscr{A}] \bm{\phi}(x) ) (\mathscr{D}_{\mu}[\mathscr{A}] \bm{\phi}(x) )
\} ,
\end{align}
where the Yang-Mills field $\mathscr{A}_{\mu}(x)=\mathscr{A}_{\mu}^{A}%
(x)T_{A}$ $(A=1,2,3)$ obey the gauge transformation:
\begin{align}
\mathscr{A}_{\mu}(x)  &  \to U(x) \mathscr{A}_{\mu}(x) U^{-1}(x) + ig^{-1}
U(x) \partial_{\mu}U^{-1}(x) , \quad U(x) \in G=SU(2) , \label{gauge-transf1}%
\end{align}
and the scalar field $\bm{\phi}(x)=\phi^{A}(x)T_{A}$ has the fixed radial
length (modulus) $v>0$:
\begin{align}
\bm{\phi}(x) \cdot\bm{\phi}(x) \equiv2\mathrm{tr} \{ \bm{\phi}(x) \bm{\phi}(x)
\} = \bm{\phi}^{A}(x) \bm{\phi}^{A}(x) = v^{2} , \label{SU2-YMH-1-constraint}%
\end{align}
and transforms according to the adjoint representation under the gauge
transformation:
\begin{align}
\bm{\phi}(x) \to U(x) \bm{\phi}(x) U^{-1}(x) , \quad U(x) \in G=SU(2) ,
\label{gauge-transf2}%
\end{align}
Hence there is no potential term. Therefore, the Higgs particle degrees of
freedom are eliminated.
The covariant derivative $\mathscr{D}_{\mu}[\mathscr{A}] := \partial_{\mu} -
ig[ \mathscr{A}_{\mu}, \cdot]$ transforms according to the adjoint
representation under the gauge transformation: $\mathscr{D}_{\mu}[\mathscr{A}]
\to U(x) \mathscr{D}_{\mu}[\mathscr{A}] U^{-1}(x)$. Notice that the constraint
(\ref{SU2-YMH-1-constraint}) is a gauge-invariant condition. Therefore, this
gauge-scalar model has the $SU(2)$ gauge invariance.

First, we recall the \textit{conventional description for the BEH mechanism}.
Suppose that the scalar field $\bm{\phi}(x)$ acquires a non-vanishing vacuum
expectation value (VEV): $\langle\bm{\phi}(x)\rangle=\langle\bm{\phi}\rangle
=\langle\bm{\phi}^{A}\rangle T_{A}$. Then the covariant derivative of the
scalar field is expanded as $\mathscr{D}_{\mu}%
[\mathscr{A}]\bm{\phi}(x):=\partial_{\mu}\bm{\phi}(x)-ig[\mathscr{A}_{\mu
}(x),\bm{\phi}(x)]
\rightarrow$ $
-ig[\mathscr{A}_{\mu}(x),\langle\bm{\phi}\rangle]+\cdots.$ Consequently, the
kinetic term of the scalar field reads \newline$
\mathrm{tr}\{(\mathscr{D}^{\mu}[\mathscr{A}]\bm{\phi}(x))(\mathscr{D}_{\mu
}[\mathscr{A}]\bm{\phi}(x))\}$ $\rightarrow$ $
-g^{2}\mathrm{tr}_{G}\{[T_{A},\langle\bm{\phi}\rangle][T_{B},\langle
\bm{\phi}\rangle]\}\mathscr{A}^{\mu A}(x)\mathscr{A}_{\mu}^{B}(x)+\cdots.
$
If the \textit{non-vanishing VEV $\langle\bm{\phi}\rangle=\langle
\bm{\phi}^{A}\rangle T_{A}$ of the scalar field $\bm\phi$} is chosen to a
specific direction, e.g., $\langle\bm{\phi}\rangle_{\infty}=\langle
\bm{\phi}^{3}\rangle T_{3}=vT_{3}$, [\textit{unitary gauge}] uniformly over
the spacetime, then the original local continuous gauge symmetry $G=SU(2)$ is
spontaneously broken to a subgroup $H=U(1)$.
Thus the kinetic term of the scalar field generates the mass term of the gauge
field:
\[
\mathscr{L}_{\mathrm{kin}}=\mathrm{tr}\{(\mathscr{D}^{\mu}%
[\mathscr{A}]\bm{\phi}(x))(\mathscr{D}_{\mu}%
[\mathscr{A}]\bm{\phi}(x))\}\rightarrow\frac{1}{2}(gv)^{2}(\mathscr{A}^{\mu
1}\mathscr{A}_{\mu}^{1}+\mathscr{A}^{\mu2}\mathscr{A}_{\mu}^{2}),\quad
v:=\langle\bm{\phi}^{3}\rangle.
\]
Thus we find that the off-diagonal gluons $\mathscr{A}_{\mu}^{1}%
,\mathscr{A}_{\mu}^{2}$
acquire the same mass $M_{W}:=gv=g\langle\bm{\phi}\rangle_{\infty}$, while the
diagonal gluon $\mathscr{A}_{\mu}^{3}$ remains massless.
\noindent This description of the BEH mechanism depends on the specific gauge
and is not gauge independent. Indeed, VEV $\langle\bm{\phi}\rangle_{\infty}$
is not a gauge invariant quantity.

Next, we explain a
\textit{gauge-independent description for the BEH mechanism} which does not
rely on the SSB \cite{Kondo16}.
We define a composite vector field $\mathscr{W}_{\mu}(x)$ which consists of
$\mathscr{A}_{\mu}(x)$ and $\bm{\phi}(x)$:
\begin{align}
\mathscr{W}_{\mu}(x) := -ig^{-1} [\hat{\bm{\phi}}(x), \mathscr{D}_{\mu
}[\mathscr{A}]\hat{\bm{\phi}}(x) ] , \quad\hat{\bm{\phi}}(x):=\bm{\phi}(x)/v .
\label{W-def1}%
\end{align}
We find that the kinetic term of $\bm{\phi}$ is identical to the
\textit{``mass term'' of the vector field $\mathscr{W}_{\mu}$}:
\begin{align}
\mathscr{L}_{\mathrm{kin}} = \mathrm{tr} \{ ( \mathscr{D}^{\mu}[\mathscr{A}]
\bm{\phi}(x) ) (\mathscr{D}_{\mu}[\mathscr{A}] \bm{\phi}(x) ) \} = \frac{1}{2
} M_{W}^{2} \mathscr{W}^{\mu}(x) \cdot\mathscr{W}_{\mu}(x) , \quad M_{W} := gv
, \label{W-mass}%
\end{align}
as far as the constraint ($\hat{\bm{\phi}}(x) \cdot\hat{\bm{\phi}}(x) = 1$) is
satisfied.
This \textit{``mass term'' of $\mathscr{W}_{\mu}$ is gauge invariant}, since
$\mathscr{W}_{\mu}$ obeys the adjoint gauge transformation:
\begin{align}
\mathscr{W}_{\mu}(x) \to U(x) \mathscr{W}_{\mu}(x) U^{-1}(x) .
\end{align}
The \textit{$\mathscr{W}_{\mu}$ gives a gauge-independent definition of the
massive gluon mode in the operator level.} \textit{The massive mode
$\mathscr{W}_{\mu}$ can be described without breaking the original gauge
symmetry.}
Despite its appearance of $\mathscr{W}_{\mu}$,
the independent internal degrees of freedom in $\mathscr{W}_{\mu
}=(\mathscr{W}_{\mu}^{A})$ is equal to $\mathrm{dim}(G/H)=2$, since
$\mathscr{W}_{\mu}(x) \cdot\hat{\bm{\phi}}(x)=0$.
Notice that this is a gauge-invariant statement.
Thus, \textit{$\mathscr{W}_{\mu}(x)$ represent the massive modes corresponding
to the coset space $G/H$ components} as expected.

How is this description related to the conventional one? In fact, by taking
the unitary gauge $\hat{\bm{\phi}}^{A}(x) \to\hat{\bm{\phi}}_{\infty}^{A}$,
$\mathscr{W}_{\mu}$ reduces to $\mathscr{W}_{\mu}(x) \to-ig^{-1}
[\hat{\bm{\phi}}_{\infty}, \mathscr{D}_{\mu}[\mathscr{A}]\hat{\bm{\phi}}%
_{\infty} ]
=
[\hat{\bm{\phi}}_{\infty}, [ \hat{\bm{\phi}}_{\infty}, \mathscr{A}_{\mu}(x)
]]
= \mathscr{A}_{\mu}(x) - (\mathscr{A}_{\mu}(x) \cdot\hat{\bm{\phi}}_{\infty
})\hat{\bm{\phi}}_{\infty} . $ Then $\mathscr{W}_{\mu}$ agrees with the
off-diagonal components $\mathscr{A}_{\mu}^{a}(x)$ for the specific choice
$\hat{\bm{\phi}}_{\infty}^{A}=\delta^{A3}$.
Therefore, $\mathscr{A}_{\mu}$ is separated into two pieces $\mathscr{V}_{\mu
}$ and $\mathscr{W}_{\mu}$:
\begin{align}
\mathscr{A}_{\mu}(x) = \mathscr{V}_{\mu}(x) + \mathscr{W}_{\mu}(x) ,
\quad\mathscr{W}_{\mu}(x) := -ig^{-1} [\hat{\bm{\phi}}(x), \mathscr{D}_{\mu
}[\mathscr{A}]\hat{\bm{\phi}}(x) ] . \label{A-decomposiion}%
\end{align}
We find that $\mathscr{V}_{\mu}$ is constructed from $\mathscr{A}_{\mu}$ and
$\bm{\phi}$ as
\begin{align}
\mathscr{V}_{\mu}(x) = c_{\mu}(x) \hat{\bm{\phi}}(x) + ig^{-1} [
\hat{\bm{\phi}}(x), \partial_{\mu}\hat{\bm{\phi}}(x) ] , \quad c_{\mu}(x) :=
\mathscr{A}_{\mu}(x) \cdot\hat{\bm{\phi}}(x) , \label{V-def1}%
\end{align}
and by definition transforms under the gauge transformation just like
$\mathscr{A}_{\mu}$:
\begin{align}
\mathscr{V}_{\mu}(x) \to U(x) \mathscr{V}_{\mu}(x) U^{-1}(x) + ig^{-1} U(x)
\partial_{\mu}U^{-1}(x) .
\end{align}
In the unitary gauge $\hat{\bm{\phi}}^{A}(x) \to\hat{\bm{\phi}}_{\infty}%
^{A}=\delta^{A3}$, $\mathscr{V}_{\mu}$ agrees with the diagonal component
$\mathscr{A}_{\mu}^{3}(x)$.

\section{Complementary gauge-scalar model for the Yang-Mills theory}

In the gauge-scalar model, $\mathscr{A}_{\mu}(x)$ and $\bm{\phi}(x)$ are
independent field variables.
However, the Yang-Mills theory should be described by the Yang-Mills field
$\mathscr{A}_{\mu}(x)$ alone and hence $\bm{\phi}$ must be supplied as a
composite field made from the gauge field $\mathscr{A}_{\mu}(x)$.
This is achieved by imposing the constraint which we call the
\textit{reduction condition}.
We choose e.g.,
\begin{align}
\bm{\chi}(x) := [\hat{\bm{\phi}}(x) , \mathscr{D}_{\mu}[\mathscr{A}]
\mathscr{D}_{\mu}[\mathscr{A}] \hat{\bm{\phi}}(x) ] = \bm{0}
\Longleftrightarrow\mathscr{D}_{\mu}[\mathscr{V}] \mathscr{W}_{\mu}(x) = 0 .
\label{reduction-mYM2}%
\end{align}
This condition is gauge covariant,
$
\bm{\chi}(x) \to U(x) \bm{\chi}(x) U^{-1}(x) $.
\textit{The reduction condition plays the role of eliminating the extra
degrees of freedom introduced by the radially fixed adjoint scalar field into
the Yang-Mills theory},
since $\bm{\chi}$ represents two conditions due to $\bm{\chi}(x) \cdot
\hat{\bm{\phi}}(x) = 0 . $
The ``complementary'' gauge-scalar model is defined by taking into account the
Faddeev-Popov determinant $\widetilde{\Delta}^{\mathrm{red}}$ associated with
the reduction condition $\bm{\chi}=0$ as
\begin{align}
\tilde Z_{\mathrm{RF}} = \int\mathcal{D} \mathscr{A} \mathcal{D}
\hat{\bm{\phi}} \ \delta\left(  \bm{\chi} \right)  \Delta^{\mathrm{red}}
e^{-S_{\mathrm{YM}}[\mathscr{A}]-S_{\mathrm{kin}}[\mathscr{A},v\hat
{\bm{\phi}}]} . \label{path_integral}%
\end{align}
[Without $\delta\left(  \bm{\chi} \right)  $ and $\Delta^{\mathrm{red}}$, this
model is the usual gauge-scalar model with a radially fixed scalar field.] We
perform change of variables from the original variables to the new variables:
\newline$\{ \mathscr{A}_{\mu}^{A} (x) , \hat{\bm{\phi}}^{a} (x) \} \to\{
c_{\mu} (x) , \mathscr{W}_{\nu}^{B} (x) , \hat{\bm{\phi}}^{b} (x) \} .
$ Then we have
\begin{align}
\tilde Z_{\mathrm{RF}} =  &  \int\mathcal{D} c \mathcal{D} \mathscr{W}
\mathcal{D} \hat{\bm{\phi}} \ J \delta\left(  \widetilde{\bm{\chi}} \right)
\widetilde{\Delta}^{\mathrm{red}} e^{-S_{\mathrm{YM}}%
[\mathscr{V}+\mathscr{W}]-iS_{\mathrm{m}}[\mathscr{W}] } , \ S_{\mathrm{m}%
}[\mathscr{W}] := \int d^{D}x \frac{1}{2} M_{\mathscr{W}}^{2} \mathscr{W}_{\mu
} \cdot\mathscr{W}_{\mu} , \label{path_integral2}%
\end{align}
We can reproduce the well-known preceding cases by taking the special limit or
choosing the gauge. For instance, by taking the unitary gauge, $\bm{\phi}^{A}%
(x) = v \hat{\bm{\phi}}^{A}(x) , \ \hat{\bm{\phi}}^{A}(x) \to\delta^{A3} , $
the new variables reduce to $c_{\mu}=\mathscr{A}_{\mu}\cdot\hat{\bm{\phi}} \to
A_{\mu}^{3}, \ \mathscr{V}_{\mu}^{A} \to A_{\mu}^{3} \delta^{A3} ,
\ \mathscr{W}_{\mu}^{A} \to A_{\mu}^{a} \delta^{Aa} , $ which means
\begin{align}
\tilde Z_{\mathrm{RF}}  &  \to\int\mathcal{D}A^{3} \mathcal{D}A^{a}
\delta\left(  \mathscr{D}^{\mu}[A^{3}]A_{\mu}^{a} \right)  \Delta
_{\mathrm{FP}} e^{-S_{\mathrm{YM}}[A^{a}+A^{3}]- S_{\mathrm{m}}[A^{a}]} ,
\ S_{\mathrm{m}}[A^{a}] := \int d^{D}x \frac{1}{2} M_{\mathscr{W}}^{2} A_{\mu
}^{a} A_{\mu}^{a} .
\end{align}
In the limit, the gauge-scalar model with the radially fixed adjoint scalar
field is reduced to the Yang-Mills theory with the gauge-fixing term of the
Maximal Abelian gauge $\mathscr{D}^{\mu}[A^{3}]A_{\mu}^{a}=0$ and the
associated Faddeev-Popov determinant $\Delta_{\mathrm{FP}}$, plus a mass term
$S_{\mathrm{m}}[A^{a}]$ for the off-diagonal gluons.
In other words, \textit{the pure Yang-Mills theory in the MA gauge with the
off-diagonal gluon mass term has the gauge-invariant extension which is
identical to the gauge-scalar model with the radially-fixed adjoint scalar
field subject to the reduction condition}, which we call the ``complementary''
gauge-scalar model.

The field strength $\mathscr{F}_{\mu\nu}[\mathscr{V}](x)$
of $\mathscr{V}_{\mu}(x)$ is shown to be proportional to $\hat{\bm{\phi}}%
(x)$:
$\mathscr{F}_{\mu\nu}[\mathscr{V}](x)
= \hat{\bm{\phi}}(x) \{ \partial_{\mu}c_{\nu}(x) - \partial_{\nu}c_{\mu}(x) +
H_{\mu\nu}(x) \} , \
H_{\mu\nu}(x) := ig^{-1} \hat{\bm{\phi}}(x) \cdot[\partial_{\mu}%
\hat{\bm{\phi}}(x) , \partial_{\nu}\hat{\bm{\phi}}(x) ] .
$ Then we can introduce the Abelian-like \textit{$SU(2)$ gauge-invariant field
strength} $f_{\mu\nu}$ by $
f_{\mu\nu}(x) := \hat{\bm{\phi}}(x) \cdot\mathscr{F}_{\mu\nu}[\mathscr{V}](x)
= \partial_{\mu}c_{\nu}(x) - \partial_{\nu}c_{\mu}(x) + H_{\mu\nu}(x) . $
In the low-energy $E \ll M_{W}$ or the long-distance $r \gg M_{W}^{-1}$
region, the massive modes $\mathscr{W}_{\mu}(x)$ become negligible and the
restricted (residual) fields
become dominant the low-energy region is described by the \textit{restricted
Yang-Mills theory}: $\mathscr{L}_{\mathrm{YM}}^{\mathrm{rest}} = - \frac{1}{4}
\mathscr{F}^{\mu\nu}[\mathscr{V}] \cdot\mathscr{F}_{\mu\nu}[\mathscr{V}] = -
\frac{1}{4} f^{\mu\nu} f_{\mu\nu} .
$
Thus, the \textit{``Abelian'' dominance in quark confinement is understood as
a consequence of the BEH mechanism for the ``complementary'' gauge-scalar
model in the gauge-invariant way.}
Moreover, \textit{if the fields $\mathscr{A}$ and $\bm{\phi}$ are a set of
``solutions'' of the field equations for the gauge-scalar model with a
radially fixed scalar field, they are automatically field configurations
satisfying the reduction condition for the pure Yang-Mills theory}%
\cite{Kondo16}.

\section{Massive Yang-Mills theory on the lattice}

We now discuss the numerical simulations for the proposed massive Yang-Mills
theory on the lattice. By taking into account the reduction condition in the
complementary gauge-scalar model, the gauge-invariant mass term is
introduced:
\begin{align}
&  Z_{L}=\int\mathcal{D}[U]\mathcal{D}[\mathbf{n}]\mathbf{\delta
}(\mathbf{n-\hat{n}})e^{-\beta S_{g}-\gamma S_{m}},\\
&  S_{g}[U]:=\sum_{x}\sum_{\mu>\nu}2{Re}\mathrm{tr}\left(  \mathbf{1}%
-U_{x,\mu}U_{x+\mu,\nu}U_{x+\nu,\mu}^{\dag}U_{x,\nu}^{\dag}\right)  ,\\
&  S_{m}[U,\mathbf{n}]:=\sum_{x,\mu}\mathrm{tr}\left(  (D_{\mu}^{\epsilon
}[U]\mathbf{n}_{x})^{\dag}(D_{\mu}^{\epsilon}[U]\mathbf{n}_{x})\right)  ,\quad
D_{\mu}^{\epsilon}[U]\mathbf{n}_{x}:=U_{x,\mu}\mathbf{n}_{x+\mu}%
-\mathbf{n}_{x}U_{x,\mu},
\end{align}
where $U_{x,\mu}$ $\in SU(2)$ is the link variable, $\mathbf{n}%
\mathcal{\mathbf{=}}n_{A}T^{A}$ $\in su(2)$ is the color field (scalar field
$\bm{\phi}$) with $\mathbf{n\cdot n}=1$, and $D_{\mu}^{\epsilon}%
[U]\mathbf{n}_{x}$ is the covariant derivative. Here $\mathbf{\delta
}(\mathbf{n-\hat{n}})$ represents the reduction condition in the complementary
gauge-scalar model, and $\mathbf{\hat{n}}$ is the solution of the reduction
condition for a given gauge configuration, which is obtained by minimizing the
functional:
\begin{equation}
F_{\text{red}}[\mathbf{n;}U]:=\sum_{x,\mu}\mathrm{tr}\left(  (D_{\mu
}^{\epsilon}[U]\mathbf{n}_{x})^{\dag}(D_{\mu}^{\epsilon}[U]\mathbf{n}%
_{x})\right)  .\label{eq:reduction}%
\end{equation}
Now, we perform the numerical simulation to generate the gauge field
configuration:
\begin{equation}
\rho\lbrack U,\mathbf{n}]:=\frac{\mathbf{\delta}(\mathbf{n-\hat{n}})e^{-\beta
S_{g}-\gamma S_{m}}}{Z_{L}}\text{ }.
\end{equation}
In absence of the reduction condition, this model is the usual gauge-scalar
model with a radially fixed scalar field.
If $\gamma=0$, the model reduces to the usual Yang-Mills theory with the
standard Wilson action.
In the massive Yang-Mills theory, $U_{x,\mu}$ and $\mathbf{n}$ are no longer
independent field variables.
For the theory to be described by $U_{x,\mu}$ alone, the color field
$\mathbf{n}$ must be supplied as a composite field made from $U_{x,\mu}$. This
is achieved by imposing the reduction condition.
Thus, the gauge field configurations must be updated by solving the reduction
condition simultaneously.
As the first step, we investigate the region, $\gamma\sim0$ by using the
reweighting technique.

We have performed the numerical simulation on the $32^{4}$ lattice using the
standard Wilson action at $\beta=2.5$ and $\gamma=0$ with over-relaxation
algorithm.
After 80000 sweeps thermalization, we have generated 4000 configurations every
400 sweeps.
To obtain the color field configurations, we have solved the reduction
condition for each gauge configuration by minimizing the functional
(\ref{eq:reduction}). The color field $\mathbf{\hat{n}}$ is obtained as a
function of the gauge configuration: $\mathbf{\hat{n}}=\mathbf{\hat{n}%
}[\mathbf{U}]$.
The observable $\mathcal{O}$ is measured by using the reweighting method as
\begin{equation}
\left\langle \mathcal{O}\right\rangle :=\frac{\sum\mathcal{O}[U,\mathbf{\hat
{n}}]e^{-\gamma S_{m}[U,\mathbf{\hat{n}}]}}{\sum e^{-\gamma S_{m}%
[U,\mathbf{\hat{n}}]}}.
\end{equation}
We study the Wilson loop $W[R,T]$ of size $R\times T$ and the
\textquotedblleft mass term\textquotedblright\ $M$ defined by
\begin{equation}
M:=\frac{1}{N_{\text{site}}}\sum_{x,\mu}\mathrm{tr}\left(  (D_{\mu}^{\epsilon
}[U]\mathbf{n}_{x})^{\dag}(D_{\mu}^{\epsilon}[U]\mathbf{n}_{x})\right)
.\label{eq;M}%
\end{equation}

In what follows, figures show the preliminary results of numerical
simulations, where no error bars are plotted because they are very large for
finite $\gamma$.%

\begin{figure}[hbt] \centering
\includegraphics[height=40mm, clip, viewport=50 200 560 620 ]
{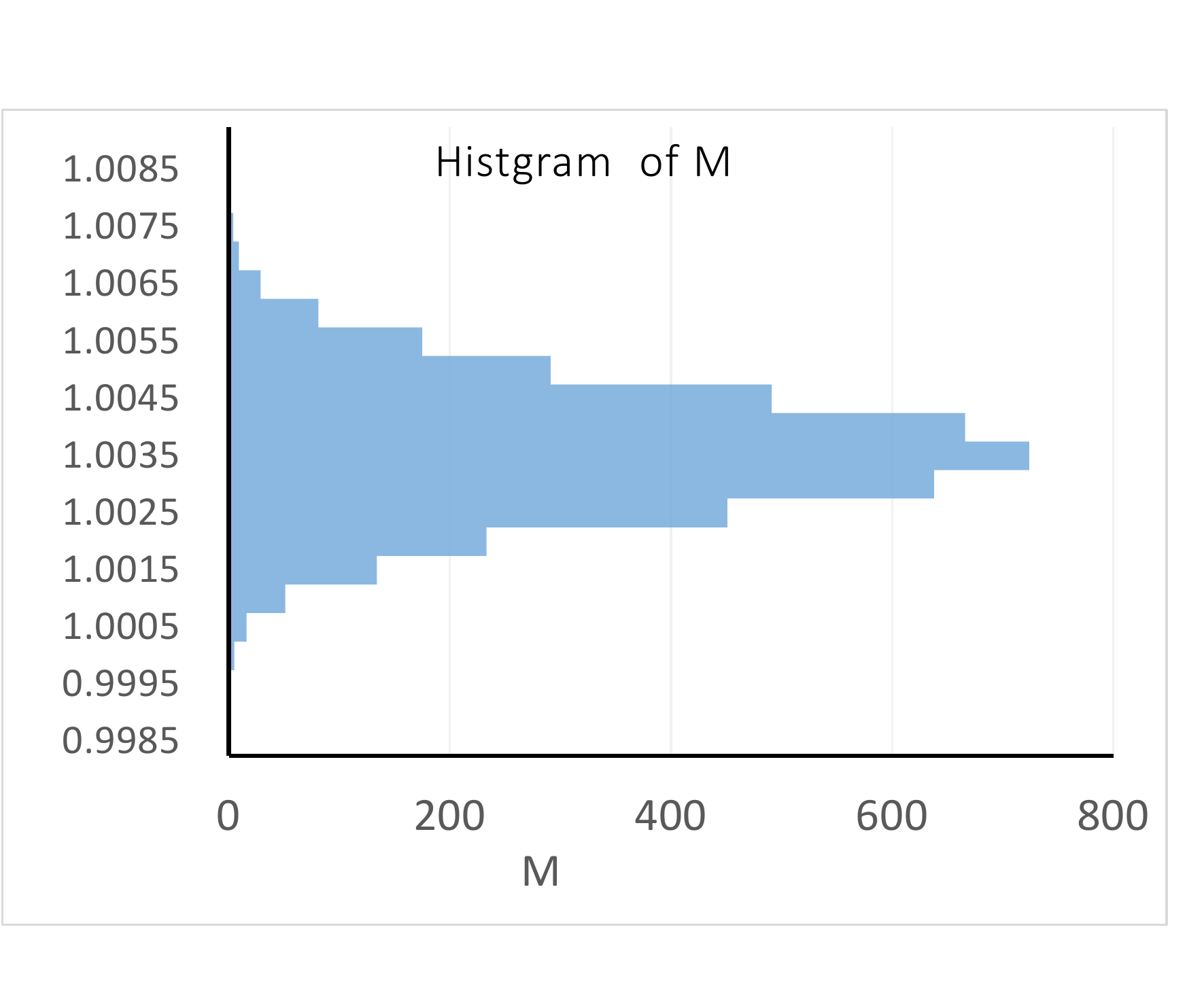}%
\includegraphics[height=40mm, clip, viewport=100 0 700 550 ]
{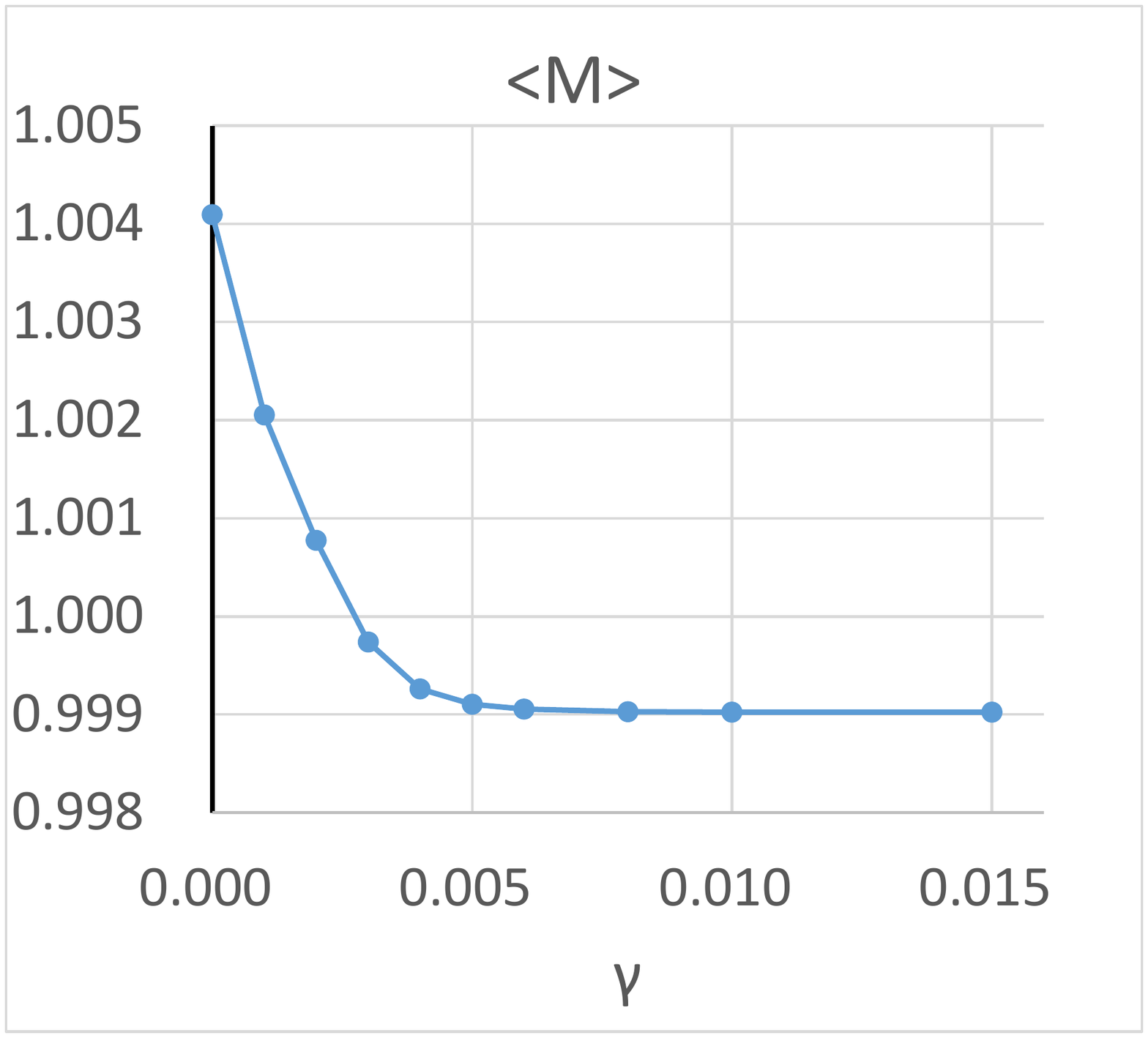}%
\includegraphics[height=40mm, clip,  viewport = 0 0 700 550 ]
{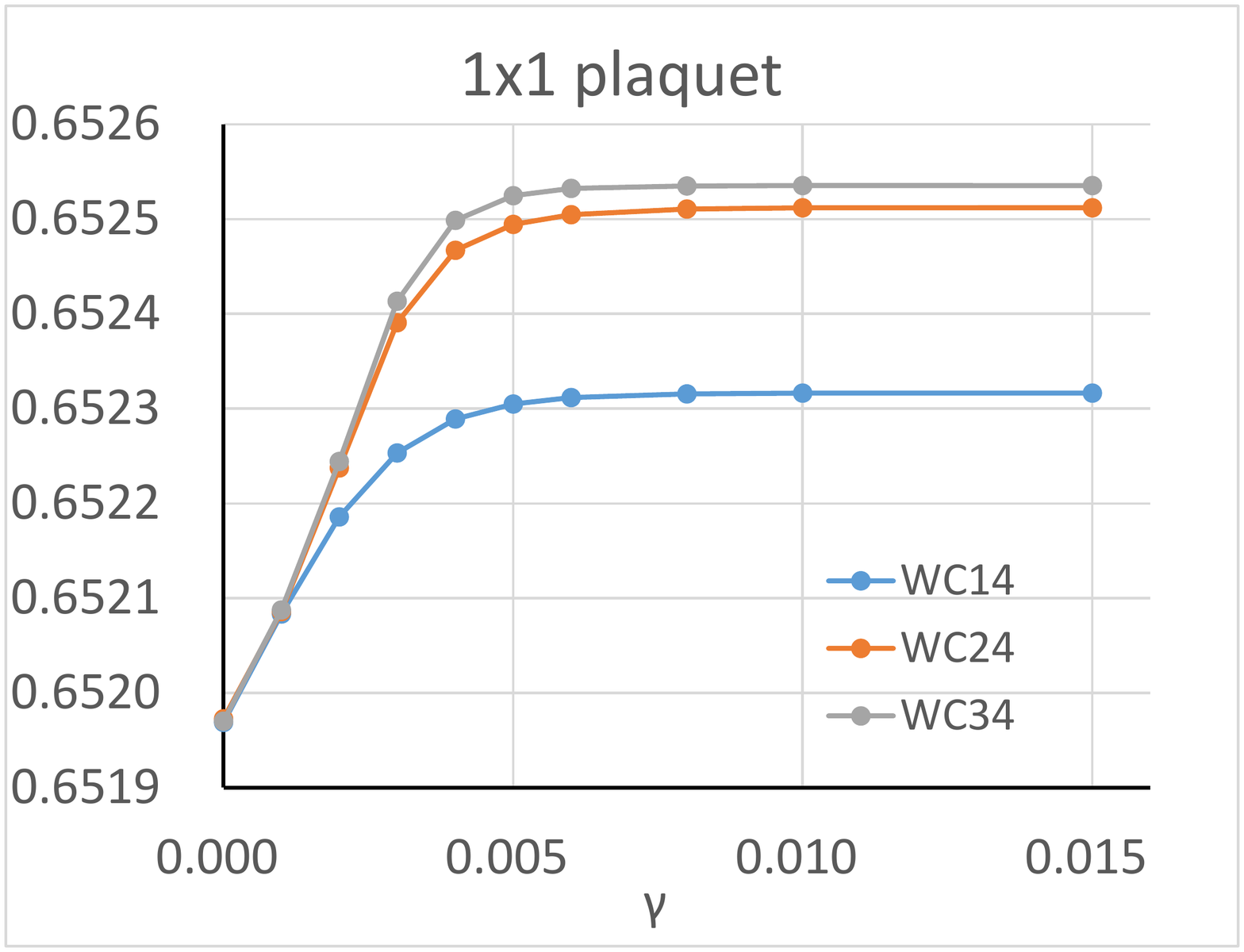}%
\caption{
(Left) the histgram of the mass term $M$,
(Middle) the measurement of the $\left\langle M \right\rangle$,
(Right) the measurement of the $1 \times1 $ plaquette
on the plain of $X-T$ (WC14),  $Y-T$ (WC24),  $Z-T$ (WC34) .
}%
\label{Fig:histgram}%
\end{figure}%

The left panel of Fig.\ref{Fig:histgram}~ shows the histogram of the value $M$
in (\ref{eq;M}) for using configurations \ The middle panel of
Fig.\ref{Fig:histgram} show the expectation value of $M$ for various values of
$\gamma$. The expectation value $\left\langle M\right\rangle $ is the
decreasing function of $\gamma,$ and the function becomes flat, when $\gamma$
is larger than $\gamma_{0}$ $(\simeq0.004)$ \ This could be due to the fact
that there is no overlap between the reweighted configuration distribution and
the original configuration distribution. The right panel of
Fig.\ref{Fig:histgram} shows measurement of the plaquette for various
parameter $\gamma$ \ %

\begin{figure}[hbt] \centering
\includegraphics[height=39mm]{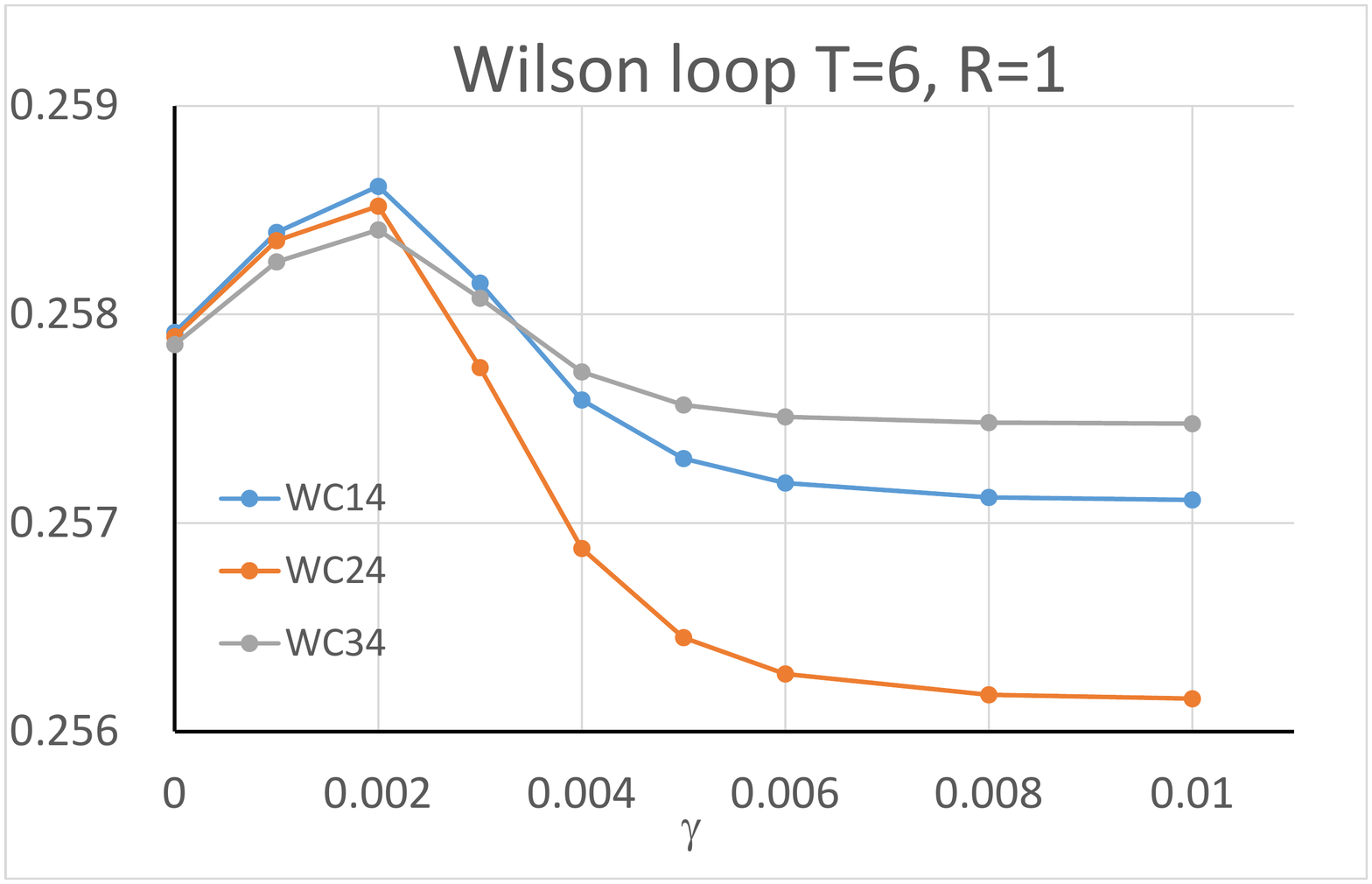}%
\includegraphics[height=39mm, clip, viewport=50 200 550 600 ]
{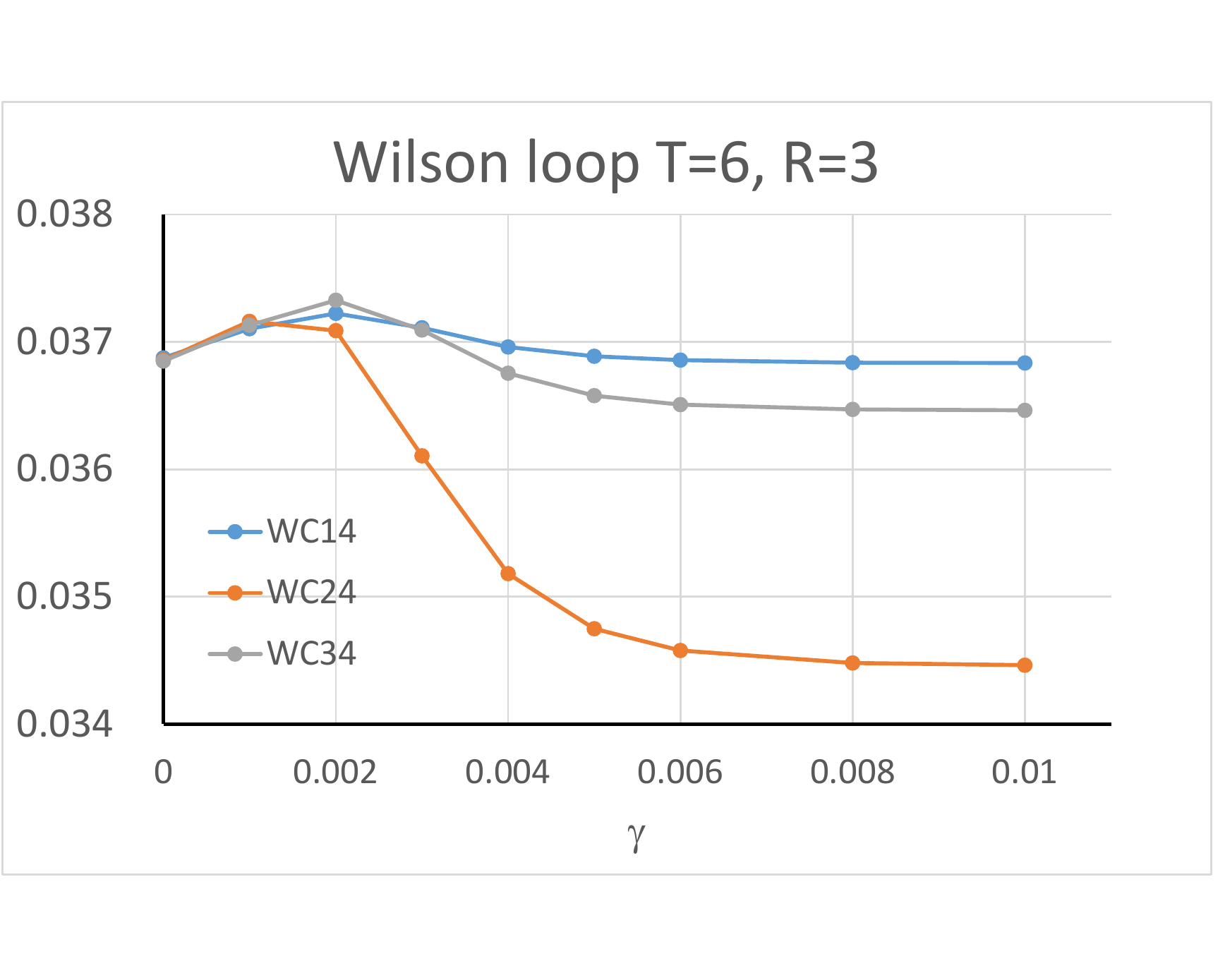}%
\includegraphics[height=39mm]{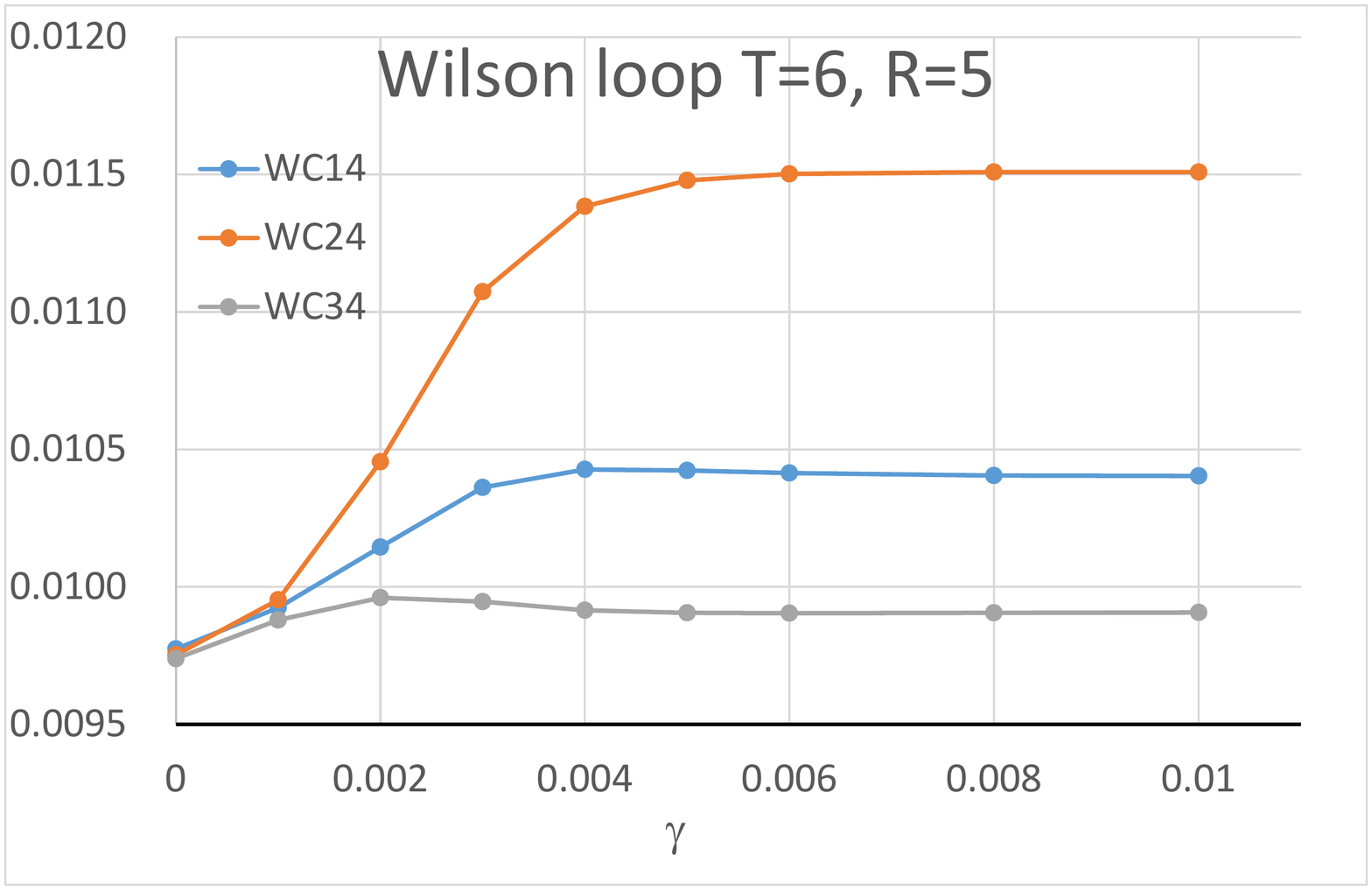}%
\caption{
Wilson loop average for X-T, Y-T, Z-T plane. (Left) R=1, T=6, (Middle) R=3, T=6, (Right) R=5, T=6.
}%
\label{Fig:Wilson-loop}%
\end{figure}%

Fig.~\ref{Fig:Wilson-loop} shows the measurement of the Wilson loop average
for various $\gamma$ \ for $X-T$, $Y-T$, $Z-T$ planes. The Wilson loop average
increases in $\gamma$ for small $\gamma$, but the behavior changes when
$\gamma$ is larger than $\gamma_{1}$ ($\simeq0.002$)$.$%

\begin{figure}[hbt] \centering
\includegraphics[height=55mm]{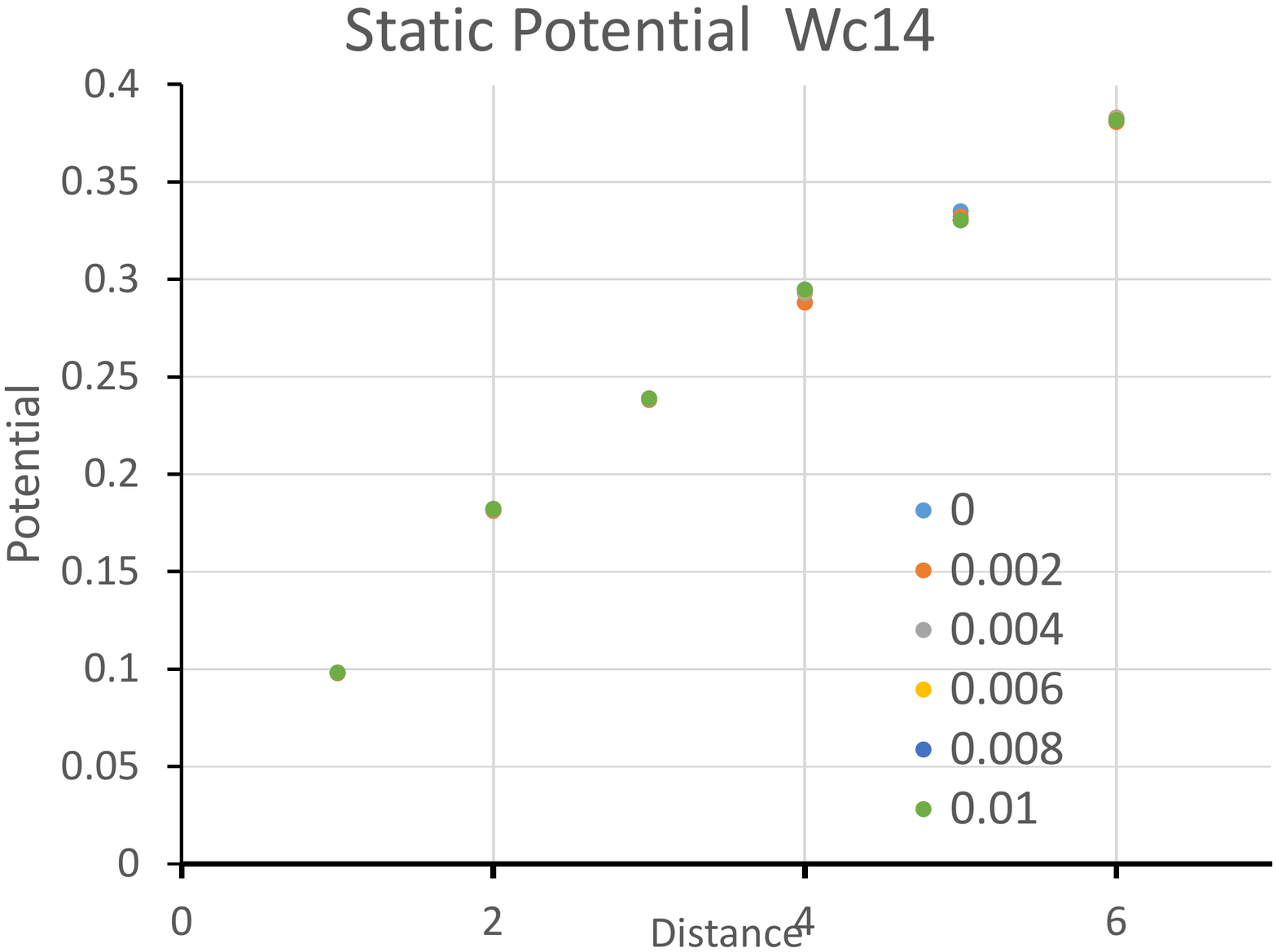}%
\includegraphics[height=55mm]{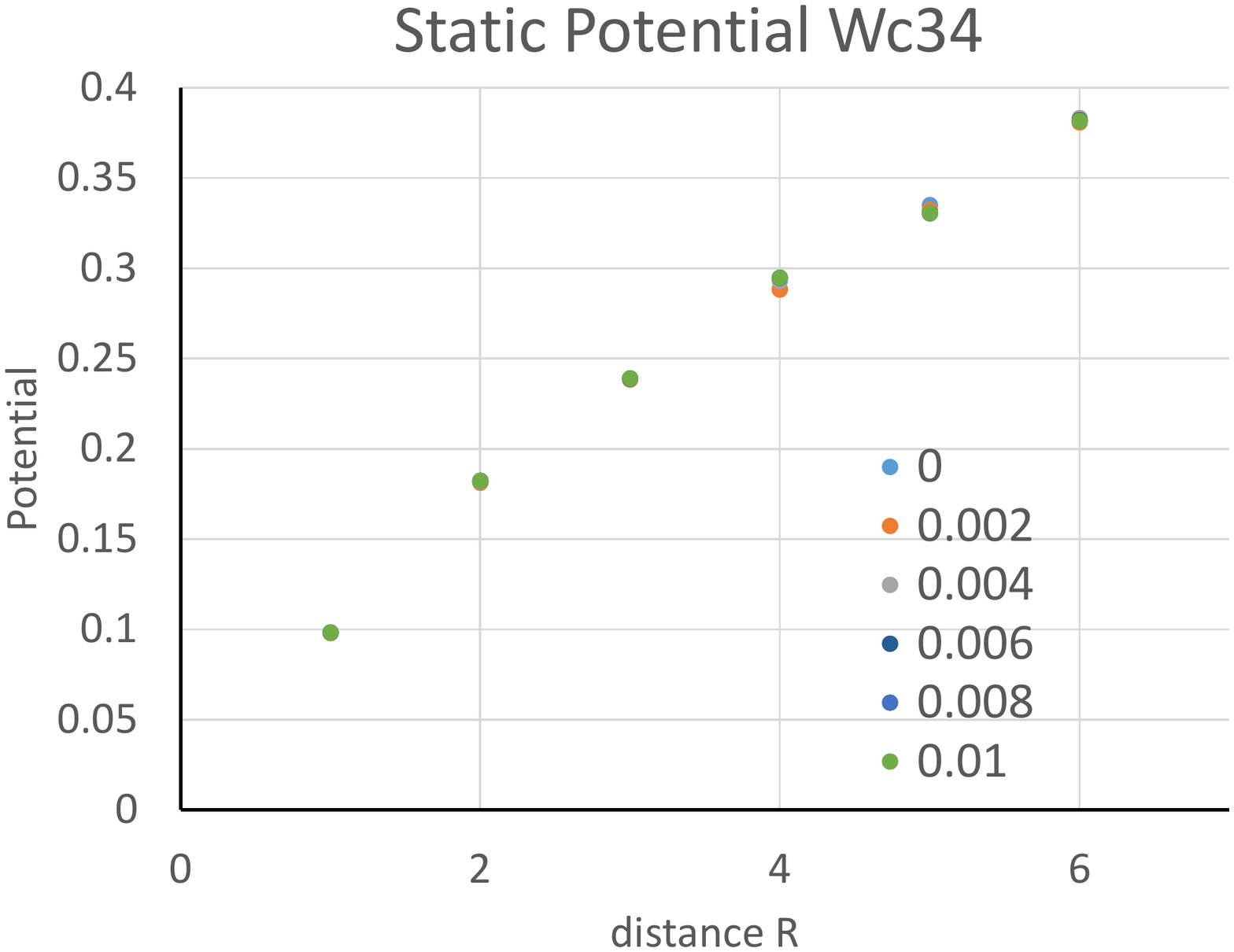}%
\caption{
The static potentials for various values of $\gamma$.
(Left) for the X--T plane, (Right) for the Z-T plane.
}%
\label{Fig:potential}%
\end{figure}%

Finally, Fig.~\ref{Fig:potential} show results of the static potential for
various values of $\gamma$.


\section{Summary and Outlook}

In this talk, we have given the first lattice calculation of Yang-Mills theory
with a gauge-invariant gluon mass term in the region of a small mass parameter
$\gamma$ by using the reweighting technique. It has been found that the
simulation can only be performed in a very small region of $\gamma$, and that
the full simulation without the reweighting technique is indispensable to
investigate the whole parameter space of the gauge coupling $\beta$ and the
mass parameter $\gamma$. These are very preliminary results to be improved in
due course.

\subsubsection*{Acknowledgement}

This work was supported by Grant-in-Aid for Scientific Research, JSPS KAKENHI
Grant Number (C) No.19K03840.

\end{document}